\documentclass[12pt]{article}
\def\la{\lambda}

\newcommand{\beq}{\begin{equation}}
\newcommand{\eeq}{\end{equation}}
\newcommand{\bea}{\begin{eqnarray*}}
\newcommand{\eea}{\end{eqnarray*}}
\newcommand{\beqa}{\begin{eqnarray}}
\newcommand{\eeqa}{\end{eqnarray}}
\begin{document}

\begin{titlepage}

\renewcommand{\thefootnote}{\fnsymbol{footnote}}

\hfill\parbox{4cm}{hep-th/0511195 }
\vskip 1.5cm

\centerline{\Large\bf Deformed Type 0A Matrix Model and }
\vskip .5cm
\centerline{\Large\bf Super-Liouville Theory for Fermionic Black Holes}
\vskip 1cm

\centerline{\large Changrim Ahn$^{1}$\footnote{\tt
    ahn@ewha.ac.kr},
Chanju Kim$^{1}$\footnote{\tt cjkim@ewha.ac.kr},
Jaemo Park$^{2}$\footnote{\tt jaemo@physics.postech.ac.kr} ,}
\vskip .5cm
\centerline{\large Takao Suyama$^{3}$\footnote{\tt
    suyama@gauge.scphys.kyoto-u.ac.jp},
and Masayoshi Yamamoto$^{2}$\footnote{\tt yamamoto@dante.ewha.ac.kr}}
\vskip .5cm
\centerline{$^{1}$ Department of Physics}
\centerline{Ewha Womans University}
\centerline{Seoul 120-750, Korea}
\vskip .5cm
\centerline{$^{2}$ Department of Physics}
\centerline{Pohang University of Science and Technology}
\centerline{Pohang 790-784, Korea}
\vskip .5cm
\centerline{$^{3}$ Department of Physics}
\centerline{Kyoto University}
\centerline{Kitashirakawa, Kyoto 606-8502, Japan}
\vskip .5cm

\centerline{\bf Abstract}
\vskip .5cm
We consider a ${\hat c}=1$ model in the fermionic black hole background.
For this purpose we consider a model which contains both the $N=1$ and the $N=2$
super-Liouville interactions.
We propose that this model is dual to a recently proposed
type 0A matrix quantum mechanics model with vortex deformations.
We support our conjecture by showing that non-perturbative corrections to the free energy
computed by both the matrix model and the super-Liouville theories agree exactly
by treating the $N=2$ interaction as a small perturbation.
We also show that a two-point function on sphere calculated from the deformed
type 0A matrix model is consistent with that of the $N=2$ super-Liouville theory
when the $N=1$ interaction becomes small.
This duality between the matrix model and super-Liouville theories
leads to a conjecture for arbitrary $n$-point correlation functions
of the $N=1$ super-Liouville theory on the sphere.

\end{titlepage}

\section{Introduction}

Low dimensional (super-)string theories, minimal conformal field theories (CFTs) coupled with
world sheet quantum (super-)gravity \cite{LFT} have been actively studied
mainly because they can be toy models useful to test many interesting conjectures
which are difficult to deal with in higher dimensions.
Such dualities as holography and open/closed string duality can be quantitatively confirmed
by calculating rigorously various amplitudes from both the world sheet formulations and
their dual matrix models.
Another important application of the (super-)Liouville theories is to investigate black
hole physics.
One of crucial developments is that 2D Euclidean black hole constructed by
a coset CFT $SL(2,R)/U(1)$ \cite{Witten} is dual to the sine-Liouville theory \cite{FZZi}.

Dual matrix model for the black hole is provided by Kazakov, Kostov, and Kutasov \cite{KKK}.
These authors have investigated the $c=1$ matter CFT coupled with the Liouville theory
which is perturbed by the sine-Liouville interaction.
When the cosmological constant, the coefficient of the Liouville interaction,
is much bigger than that of the sine-Liouville interaction,
the model is basically the $c=1$ string theory and described by conventional
matrix quantum mechanics.
The opposite limit in which the sine-Liouville interaction dominates
describes the Witten's 2D Euclidean black hole.
The corresponding matrix model can be constructed by adding vortex operators to the
matrix quantum mechanics.

This conjecture is proved by computing the free energies of both theories directly.
The free energy of the deformed matrix model with infinite number of vortex deformations
is identified with the $\tau$-function of the Toda chain hierarchy.
Allowing only the sine-Liouville perturbation, the hierarchy is simplified to the
Toda equation, which can be solved exactly to find the free energy as a function of
the cosmological constant and the coefficient of the sine-Liouville interaction.
The free energy can be expressed in terms of an infinite sum of correlation functions of
the sine-Liouville interaction terms.
Using G. Moore's conjecture for the correlation functions \cite{moore}, one can
sum up the infinite terms to derive the free energy which agrees with the matrix model result.

The dual matrix model for the coupled Liouvile theory is further confirmed by
Alexandrov, Kazakov and Kutasov who computed non-perturbative corrections of both theories \cite{AKK}.
The free energy of a string theory is given by the genus expansion
\begin{equation}
F_{\rm pert}=\sum_{g=0}^{\infty}g_s^{2g-2}f_g
\end{equation}
where $g_s$ is the string coupling constant and $g$ is the genus of the Riemann surfaces.
This perturbative free energy is corrected by non-perturbative effects which are
given as a form of
\begin{equation}
F_{\rm non-pert}\approx C g_s^{f_A}e^{-f_D/g_s}.
\end{equation}
These effects are the exponentiation of the disk partition sum of the D-instanton.
To work out the non-perturbative effects in the Liuoville theory is a difficult task
because the understanding of the D-brane involves the strong coupling effect
as $g_s \sim \exp(Q \phi)$ with $\phi$ representing Liouville direction.
This problem was solved by \cite{FZZ, ZZ} where the extended (FZZT) and localized (ZZ)
branes were found.
For the Liouville theory, the leading non-perturbatve correction is given by the
fundamental ZZ-brane and consistent with the matrix model result.

In this paper we generalize these developments to the supersymmetric theories.
The matrix model for the fermionic black holes has been constructed by adding
vortex deformations to the type 0A matrix quantum mechanics \cite{JaemoSuyama}.
Because the type 0A matrix model is dual to
the ${\hat c}=1$ conformal matter coupled by the $N=1$ super-Louville field theory (SLFT)
\cite{newhat} and the fermionic black hole to the $N=2$ SLFT \cite{HoriKapustin},
it is natural to conjecture that the deformed type 0A matrix model
is dual to the $N=1$ super-Louville theory perturbed by the $N=2$ SLFT interaction.

We support our conjecture using many recent developments on the SLFT theories.
For the $N=1$ SLFT, the structure constants \cite{RS,Pog} and boundary one-point functions
\cite{FH, ARS} have been calculated.
Similar results for the $N=2$ SLFT have been obtained in
\cite{AKRS,EguSug,RibSch,ASY,McGMurVer,IPT,FNP,Hosomichi}.
Using these results, we will calculate the free energy of the coupled SLFT
and non-perturbative corrections due to the D-branes and show that they agree with the
deformed type 0A matrix model results.

This paper is organized as follows: In sect.2 we calculate the free energy and
non-perturbative corrections of the deformed type 0A matrix model based on \cite{JaemoSuyama}.
The $N=1$ SLFT perturbed by the $N=2$ SLFT interaction is
defined in sect.3.
To compare with matrix model results, we compute the non-perturbative corrections based on
the boundary one-point functions of the $N=1$ SLFT and a bulk two-point function of
the $N=2$ SLFT.
We conclude the paper with discussions and open questions in sect.4.
We provide necessary formulae of the $N=1$ and $N=2$ SLFTs in the Appendices.

\section{Deformed Type 0A Matrix Quantum Mechanics}

\subsection{Perturbative expansion of the free energy}

In this section, we set $\alpha'=1/2$.
A matrix model of the type 0A string theory in two dimensions without R-R flux is
proposed \cite{newhat}
\begin{equation}
S_{0A} = \int dt\ \mbox{Tr}\Big[ |Dt|^2-|t|^2 \Bigr].
\label{0AMM}
\end{equation}
Here $t$ is an $N\times N$ complex matrix and $Dt$ is the covariant derivative.
We will consider the case in which the time direction is compactified on $S^1$
with radius $R$, so there are
two non-trivial Wilson loops Tr$\Omega,$ Tr$\tilde{\Omega}$ for two gauge fields.
As in \cite{KKK}, one can deform the type 0A matrix model (\ref{0AMM})
as follows \cite{JaemoSuyama}:
\begin{equation}
S = S_{0A}
+\sum_{n\ne0}\Bigl[ \lambda_n\mbox{Tr}\Omega^n+\tilde{\lambda}_n\mbox{Tr}\tilde{\Omega}^n
\Bigr].
\end{equation}
When the parameters are given by
\begin{equation}
\lambda_{\pm1}=\tilde{\lambda}_{\pm1}=\lambda, \hspace{5mm}
\lambda_{n\ne\pm1}=\tilde{\lambda}_{n\ne\pm1}=0,
\label{Witten's}
\end{equation}
this corresponds to a matrix model of Witten's black hole in type 0A string theory.

When $\lambda_n=\tilde{\lambda}_n$,
the partition function of
the deformed matrix model is factorized as follows,
\begin{eqnarray}
Z_{0A}(\lambda,\mu)
&=& \sum_{N=0}^\infty e^{2\pi R\mu N}\int D\Omega D\tilde{\Omega}Dt\ e^{-S} \nonumber \\
&=& Z_+(\lambda,\mu)\cdot Z_-(\lambda,\mu).
\end{eqnarray}
Moreover, when the condition (\ref{Witten's}) holds,
the free energy of each factor defined by $F_\pm(\lambda,\mu):=\log Z_\pm(\lambda,\mu)$ satisfies
the following non-linear differential equation,
\begin{equation}
\frac14\lambda^{-1}\partial_\lambda(\lambda\partial_\lambda F_\pm(\lambda,\mu'))
+\exp\Bigl[ -\sin^2(\partial_{\mu'}F_\pm(\lambda,\mu') \Bigr] = 1,
  \label{Toda}
\end{equation}
where $\mu'=\mu/2$.
From now on, we will drop the prime.

One can solve this equation at least perturbatively, by using the free energy at $\lambda=0$
\begin{eqnarray}
F_{\pm}(\lambda=0,\mu)
&=& -\frac14\mbox{Re}\int_0^\infty\frac{dt}t\frac{e^{-it\mu}}{\sinh(t/2)\sinh(t/4R)} \nonumber \\
& & \mp\frac i4\mbox{Im}\int_0^\infty\frac{dt}t\frac{e^{-it\mu}}{\sinh(t/2)\cosh(t/4R)},
\label{bc}
\end{eqnarray}
as a boundary condition.
For the tree level free energy $F_{\pm,0}(\lambda,\mu)$
the susceptibility
\beq
\chi_{\pm,0}=\partial_\mu^2F_{\pm,0}=-{2R\over{1-R}}\log\lambda+X_{\pm,0}(y),
\eeq
satisfies the following differential equation
\begin{equation}
(y\partial_y)^2X_{\pm,0}(y)+4(1-R)^2\partial_y^2e^{-X_{\pm,0}(y)}=0,
\end{equation}
where $y=\mu/\lambda^{1/(1-R)}$.
The solution is
\begin{equation}
y=e^{-X_{\pm,0}/2R}-(2R-1)e^{(1-2R)X_{\pm,0}/2R}.
\label{Xpmzero}
\end{equation}
From this we can derive
\begin{equation}
\chi_{\pm,0}(\lambda,\mu)=-2R\log\mu+2R\log(1-s)
\label{chizerolamu}
\end{equation}
where
\begin{equation}
{s\over{(1-s)^{2-2R}}}=(2R-1)\mu^{2R-2}\lambda^2.
\end{equation}
Notice that this result is the same as that of \cite{KKK} if we rescale $R$ to $2R$.

Eq.(\ref{chizerolamu}) can be used to expand the susceptibility $\chi_{\pm,0}$.
For the large $y$ (small $\lambda$), the susceptibility becomes
\begin{equation}
\chi_{\pm,0}(\lambda,\mu)
= -2R\log\mu+2R\sum_{n=1}^\infty\frac1{n!}[(1-2R)\mu^{2R-2}\lambda^2]^n
\frac{\Gamma(n(2-2R))}{\Gamma(n(1-2R)+1)}.
\label{chizero}
\end{equation}
Integrating this twice we obtain the free energy
\begin{equation}
F_{\pm,0}(\lambda,\mu)
= -R\mu^2\log(\mu/\Lambda)+2R\mu^2\sum_{n=1}^\infty\frac1{n!}[(1-2R)\mu^{2R-2}\lambda^2]^n
\frac{\Gamma(n(2-2R)-2)}{\Gamma(n(1-2R)+1)}.
\label{Fpmzero}
\end{equation}
When $\lambda$ becomes strong, we can take a series expansion of the $N=1$ SLFT
perturbation for the underlying $N=2$ SLFT.
From Eq.(\ref{chizerolamu}) we obtain for $1/2<R<1$
\begin{eqnarray}
\chi_{\pm,0}(\lambda,\mu)
&=&-{R\over{1-R}}\log\left((2R-1)\lambda^2\right)\nonumber\\
&+&{R\over{1-R}}
\sum_{n=1}^\infty\frac1{n!}\left[-\mu\left((2R-1)\lambda^2\right)^{{1\over{2-2R}}}\right]^n
\frac{\Gamma\left({n\over{2-2R}}\right)}{\Gamma\left({n\over{2-2R}}-n+1\right)}.
\label{chizeroopp}
\end{eqnarray}

When $R=1/2$ which corresponds to the fermionic black hole, one needs to deal
with Eq.(\ref{Fpmzero}) carefully because the first two terms becomes singular
while the expansion parameter vanishes.
A careful calculation leads to \cite{JaemoSuyama}
\begin{equation}
F_{\pm,0}(\lambda,\mu)
= -{\mu^2\over{2}}\log\mu-\mu\lambda^2.
\label{freeenergyatRhalf}
\end{equation}
We will compare these results with those of the $N=2$ SLFT in sect.3.

\subsection{Non-perturbative corrections}

Since the full $F_\pm(\lambda,\mu)$, not only the perturbative contributions, is a solution of
(\ref{Toda}), one can also derive non-perturbative contributions from the equation.
Suppose that the free energy is of the following form,
\begin{equation}
F_\pm(\lambda,\mu) = F_{\pm,pert}(\lambda,\mu)+\epsilon_\pm(\lambda,\mu),
\end{equation}
where the first term of the RHS contains all perturbative contributions, and non-perturbative
contributions are included in the second term.
If we obtained $F_{\pm,pert}(\lambda,\mu)$, which is possible in principle, one could obtain
$\epsilon_\pm(\lambda,\mu)$ by solving
\begin{equation}
\frac14\lambda^{-1}\partial_\lambda(\lambda\partial_\lambda\epsilon_\pm(\lambda,\mu))
= \exp\Bigl[ -4\sin^2\Bigl( \frac{\partial_\mu}2 \Bigr)F_{\pm,pert}(\lambda,\mu) \Bigr]
\Bigl\{ 1-\exp\Bigl[ -4\sin^2\Bigl( \frac{\partial_\mu}2 \Bigr)\epsilon_\pm(\lambda,\mu) \Bigr]
\Bigr\}.
\end{equation}

One may notice from (\ref{bc}) that the leading non-perturbative term would behave like
$e^{-2\pi R\mu}$ coming from the imaginary part of $F_\pm(\lambda=0,\mu)$.
This contribution is not physical since this term would be cancelled in the total free energy
$F_{0A}(\lambda,\mu)=F_+(\lambda,\mu)+F_-(\lambda,\mu)$.
So the physically relevant term would come from the next-leading terms.
We denote the leading non-perturbative term by $\epsilon_0(\lambda, \mu)$
and the next-to-leading term by $\epsilon_1(\lambda, \mu)$.
The equations one has to solve are now,
\begin{eqnarray}
\frac14\lambda^{-1}\partial_\lambda(\lambda\partial_\lambda\epsilon_0(\lambda,\mu))
&=& 4\exp\Bigl[ -4\sin^2\Bigl( \frac{\partial_\mu}2 \Bigr)F_{\pm, pert}(\lambda,\mu) \Bigr]
    \sin^2\Bigl( \frac{\partial_\mu}2 \Bigr)\epsilon_0(\lambda,\mu),
    \label{leading} \\
\frac14\lambda^{-1}\partial_\lambda(\lambda\partial_\lambda\epsilon_1(\lambda,\mu))
&=& \exp\Bigl[ -4\sin^2\Bigl( \frac{\partial_\mu}2 \Bigr)F_{\pm,pert}(\lambda,\mu) \Bigr]
    \Bigl\{ 4\sin^2\Bigl( \frac{\partial_\mu}2 \Bigr)\epsilon_1(y,\mu) \nonumber \\
& & -8\Bigl( \sin^2\Bigl( \frac{\partial_\mu}2 \Bigr)\epsilon_0(y,\mu) \Bigr)^2  \Bigr\}.
      \label{subleading}
\end{eqnarray}
In solving (\ref{leading}) and (\ref{subleading}), we use the tree level free energy expression
$F_{\pm, 0}(\lambda, \mu)$ for $F_{\pm, pert}(\lambda, \mu)$.
By making the following ansatz for $\epsilon_0(\lambda,\mu)$
\begin{equation}
\epsilon_0(\lambda,\mu) = P_0(y,\mu)e^{-\mu f(y)},
\end{equation}
one can obtain, from (\ref{leading}), the following equation for $f(y)$,
\begin{equation}
\frac1{2(1-R)}(1-y\partial_y)g(y) = \pm e^{-X(y)/2}\sin(\partial_yg(y)),
   \label{e_I3}
\end{equation}
where $g(y)=yf(y)/2$ and $X(y)\equiv X_{\pm, 0}$.
We impose the boundary condition $f(y)\to 2\pi R$ as $y\to\infty$.
Then we obtain the solution
\begin{equation}
f(y) = 2\pi R\pm4\sin(\pi R)\lambda\mu^{R-1}+\cdots,
\end{equation}
for large $\mu$.
Note that $R$ should be less than one in order to obtain $F_{\pm,pert}$ obeying the boundary
condition.

In deriving (\ref{subleading}), we assumed that $\epsilon_1\sim(\epsilon_0)^2$.
Therefore $\epsilon_1(\lambda,\mu)$ should behave as
\begin{equation}
\epsilon_1(\lambda,\mu) = P_1(y,\mu)e^{-2\mu f(y)}.
\end{equation}
One can show that this assumption is consistent with (\ref{subleading}).
Moreover, one can obtain both $P_0(y,\mu)$ and $P_1(y,\mu)$ perturbatively in terms of
$1/\mu$.These $1/\mu$ corrections arise if we use the free energy expressions including
contributions from higher genera.
It is nontrivial that the relation $\epsilon_1 \sim \epsilon_0^2$,
which holds for $\lambda=0$,
persists for nonzero $\lambda$ since the governing equations for
$\epsilon_0, \epsilon_1$ are nonlinear.

Now we calculate $r_m$ and $\rho_m$ in the deformed matrix model.
The definitions of $r,\ \rho$ in the matrix model are
\begin{eqnarray}
r_m &=& {\partial_\mu \log \epsilon_1\over{\sqrt{|\partial_\mu^2(F_{+,0}+F_{-,0})|}}}\
\Bigg|_{\lambda=0}, \\
\rho_m &=& {\partial_\lambda \log \epsilon_1\over{
\sqrt{|\partial_\lambda^2(F_{+,0}+F_{-,0})|}}}\ \Bigg|_{\lambda=0}.
\end{eqnarray}
Corresponding definitions in the Liouville theory are given by
\begin{eqnarray}
r &=& {\partial_{\mu_L} Z_{disk}\over{\sqrt{|\partial_{\mu_L}^2 F_0|}}}
\ \Bigg|_{\lambda=0},\label{LFTrparam} \\
\rho &=& {\partial_{\lambda_L} Z_{disk}\over{\sqrt{|\partial_{\lambda_L}^2 F_0|}}}
\ \Bigg|_{\lambda_{L}=0}\label{LFTrhoparam}
\end{eqnarray}
As explained in \cite{AKK}, $\log \epsilon$ is the disk partition sum
corresponding to the D-instanton. In the definition of $r$ and $\rho$,
the proportional constants between $\mu, \lambda$ and $\mu_L, \lambda_L$ are
canceled. When $\mu_L$ and $\lambda_L$ are defined in the next
section, we just write them as $\mu$ and $\lambda$.

From the above results, one can obtain
\begin{eqnarray}
r_m &=& -\frac{2\pi \sqrt{R}}{\sqrt{\log(\Lambda/\mu)}}, \\
\rho_m &=& -4\sin(\pi R),
\end{eqnarray}
and therefore,
\begin{equation}
\frac{\rho_m}{r_m} = \frac 2{\pi}\sqrt{\frac1R\log\frac\Lambda\mu}\sin(\pi R).
\end{equation}
In order to compare the computations in the Liouville theory where $\alpha'=2$ is used
we simply replace $R$ by $\frac{R}{2}$. Thus in this convention
\begin{eqnarray}
r_m &=& -\frac{\pi \sqrt{2R}}{\sqrt{\log(\Lambda/\mu)}},
 \label{rmatrix} \\
\rho_m &=& -4\sin(\frac{\pi R}{2}).
\end{eqnarray}
and
\begin{equation}
\frac{\rho_m}{r_m} = \frac
{2}{\pi}\sqrt{\frac{2}{R}\log\frac\Lambda\mu}\sin(\frac{\pi R}{2}).
\label{ratiomatrix}
\end{equation}
We will confirm these non-perturbative corrections by using the $N=1$ SLFT
results in sect.3.

\section{Coupled model of the $N=1$ and $N=2$ SLFTs}

\subsection{$N=1$ SLFT with the winding perturbation}

The $N=1$ SLFT with additional sine-Lioville type
interaction was also considered in \cite{JaemoSuyama}
as a $N=1$ generalization of the dual relation between the coset CFT
$SL(2,R)/U(1)$ of the Euclidean 2D black hole and the Liouville theory
with the sine-Liouville interaction. It is natural to conjecture that the
type 0A matrix model considered in the previous section
is dual to the following world sheet action:
\beq
S=\int d^2zd^2\theta\left[\frac{1}{2\pi}(DX\bar{D}X+D\Phi\bar{D}\Phi)
+i\mu e^{b\Phi}+(i\lambda e^{i\frac{R}{2}\tilde{X}+(1-\frac{R}{2})\Phi}+c.c.)\right],
\label{action}
\eeq
where $c.c.$ denote complex conjugate while
$X$, $\tilde{X}=X_{L}-X_{R}$ and $\Phi$ are real scalar superfields
\beqa
&&X=x+i\theta\xi+i\bar{\theta}\bar{\xi}+i\theta\bar{\theta}G,
\nonumber\\
&&\tilde{X}=\tilde{x}+i\theta\xi-i\bar{\theta}\bar{\xi},
~~~\tilde{x}=x_L-x_R,
\nonumber\\
&&\Phi=\phi+i\theta\psi+i\bar{\theta}\bar{\psi}+i\theta\bar{\theta}F.
\nonumber
\eeqa
In terms of the component fields this action can be written as
\beqa
S&=&\int d^2z({\cal L}_{I}+{\cal L}_{II}),
\label{actioncomp}\\
{\cal L}_{I}&=&\frac{1}{2\pi}(\partial x\bar{\partial}x
+\xi\bar{\partial}\xi+\bar{\xi}\partial\bar{\xi}
+\partial\phi\bar{\partial}\phi
+\psi\bar{\partial}\psi+\bar{\psi}\partial\bar{\psi})
+i\mu\psi\bar{\psi}e^{b\phi},
\label{n1lm}\\
{\cal L}_{II}&=&i\lambda\left[c_1\xi\bar{\xi}
-ic_2(\psi\bar{\xi}-\xi\bar{\psi})+c_3\psi\bar{\psi}\right]
e^{i\frac{R}{2}\tilde{x}+(1-\frac{R}{2})\phi}+c.c.
\nonumber\\
&+&\frac{\pi}{2}(:\mu e^{b\phi}+\lambda\left(1-\frac{R}{2}\right)
e^{i\frac{R}{2}\tilde{x}+(1-\frac{R}{2})\phi}+c.c.:)^2
\label{perturbation}
\eeqa
with
\beq
c_1=\frac{R^2}{4},\quad c_2=\frac{R}{2}\left(1-\frac{R}{2}\right),
\quad c_3=\left(1-\frac{R}{2}\right)^2.
\label{defcis}
\eeq
Here we have followed conventions in \cite{ASY} where
\begin{eqnarray}
\alpha'&=&2,\quad\partial=(\partial_x-i\partial_y)/2,\qquad
\bar{\partial}=(\partial_x+i\partial_y)/2,\nonumber\\
D&=&\frac{\partial}{\partial\theta}+\theta\partial,
~~~\bar{D}=\frac{\partial}{\partial\bar{\theta}}+\bar{\theta}\bar{\partial},
~~~\int d^2zd^2\theta=\int dxdyd\bar{\theta}d\theta.
\end{eqnarray}
As pointed out before, we need to rescale $R\to R/2$ for $\alpha'=2$ convention.
So the fermionic black hole described by the matrix model with $R=1/2$ corresponds to
$R=1$ in the SLFT.

When we set $\lambda=0$, the theory is the $N=1$ SLFT coupled with
${\hat c}=1$ free CFT
where the background charge term $Q\partial^2\phi$ is given by $Q=b+{1\over{b}}$.
The central charge of the $N=1$ SLFT is
\beq
{\hat c}_L=1+2Q^2.
\label{cenchaii}
\eeq
To cancel the conformal anomaly, we should set $b=1$ or $Q=2$
so that total central charge satisfies ${\hat c}_m+{\hat c}_L=10$.

The bulk (NS) primary fields of the $N=1$ SLFT are given by
\beq
V_\alpha(z,{\overline z})=e^{\alpha\phi}
\eeq
with conformal dimension
\beq
\Delta_\alpha={1\over{2}}\alpha(Q-\alpha).
\eeq
The last term in Eq.(\ref{perturbation}) is a contact term which can be neglected
since it does not contribute in CFT calculation.
It is convenient to define a dimensionless parameter
\beq
z={\la\over{\mu^{1-{R\over{2}}}}}.
\eeq

When $z<<1$, ${\cal L}_{II}$ can be considered as a marginal perturbation
for $0<R<2$.
In this region we can compute the effect of the ${\cal L}_{II}$ interaction
on the $N=1$ SLFT D-branes perturbatively.
Another interesting region is $z>>1$ where we should treat the last term in
${\cal L}_{I}$ which is the $N=1$ SLFT interaction as a perturbation.
In particular the ${\cal L}_{II}$ becomes the $N=2$ SLFT interaction
when $R=1$ since it can be written as
\beq
{\cal L}_{II}=\frac{i\la}{2}\psi^-\bar{\psi}^-e^{{\hat b}\phi^+}+c.c.,
\label{n2liouville}
\eeq
by redefining the fields
\beq
\phi^{\pm}=\frac{1}{\sqrt{2}}(\phi\pm i\tilde{x}),
~~~\psi^-=\frac{1}{\sqrt{2}}(\psi+i\xi),
~~~\bar{\psi}^-=\frac{1}{\sqrt{2}}(\bar{\psi}-i\bar{\xi}).
\nonumber
\eeq
Here ${\hat b}$, the $N=2$ SLFT coupling constant, becomes ${\hat b}=1/\sqrt{2}$.
It is easy to check that the $Q\partial^2\phi$ term
with $Q=2$ now acts as the background charge of the $N=2$ SLFT.
This means that $z>>1$ limit of the total theory can be interpreted as $N=2$ SLFT
perturbed by the $N=1$ SLFT interaction.
In particular the action with $\mu=0$ gives the $N=2$ SLFT which describes the
fermionic black hole, the super $SL(2,R)/U(1)$ coset CFT.

\subsection{Non-perturbative corrections}

Now we compute the non-perturbative corrections in the small $\la$ or small $z$ limit.
From Eq.(\ref{LFTrparam}), the parameter $r$ is given by
\beq
r=\frac{\partial_{\mu}Z_{disk}}{\sqrt{-\partial_{\mu}^2F_0}}
=\frac{iZ_{\hat{c}=1}\langle\psi\bar{\psi}e^{\phi}\rangle_{disk}}
{\sqrt{\langle\psi\bar{\psi}e^{\phi}\psi\bar{\psi}e^{\phi}\rangle_{sphere}}}.
\label{r1}
\eeq
Here $Z_{\hat{c}=1}$ is the partition function of the ${\hat c}=1$ matter CFT on the disk
with Neumann boundary condition.

The parameter $r$ is given by a disk one-point function and a bulk two-point function.
These correlation functions have been computed using both conformal and
modular bootstrap methods in \cite{RS,Pog,FH,ARS}.
Since the cosmological constant operator is a superconformal
descendant state of $e^{b\phi}$ with $b=1$,
the correlation functions of this operator are related to those of $e^{b\phi}$ by \cite{FH}
\beq
\langle \psi{\bar\psi}e^{b\phi}\rangle=i\eta{1\over{b^2}}\langle e^{b\phi}\rangle,
\qquad
\langle \psi{\bar\psi}e^{b\phi}\psi{\bar\psi}e^{b\phi}\rangle
=-{1\over{b^4}}\langle e^{b\phi}e^{b\phi}\rangle.
\label{descenonept}
\eeq
Here $\eta=\pm 1$ is a discrete parameter in the boundary conditions on the fermion,
$\psi+i\eta{\bar\psi}=0$.
The one-point function $\langle e^{\alpha\phi}\rangle$ is proportional to
the boundary wave function for the ZZ-brane with $(1,1)$ boundary condition given by
\beq
\Psi^{NS}_{(1,1)}(\alpha)
=\pi\left[\mu\pi\gamma\left(\frac{bQ}{2}\right)\right]^{\frac{1}{b}(\frac{Q}{2}-\alpha)}
\left[(\alpha-\frac{Q}{2})\Gamma(b(\frac{Q}{2}-\alpha))\Gamma(\frac{1}{b}
(\frac{Q}{2}-\alpha))\right]^{-1}.
\label{one}
\eeq
Therefore the one-point function with $\alpha=b$ is given by
\beq
\langle e^{b\phi}\rangle=C\Psi^{NS}_{(1,1)}(b)
=-C\pi\left[\mu\pi\gamma\left(\frac{bQ}{2}\right)\right]^{\frac{1}{2}(\frac{1}{b^2}-1)}
\left[b\Gamma(\frac{1}{2}(1-b^2))\Gamma(\frac{1}{2}(\frac{1}{b^2}+1))\right]^{-1}
\label{oneb}
\eeq
where $C$ is a proportional constant between the one-point function and boundary
wave function.

The two-point function on the sphere $\langle e^{b\phi}e^{b\phi}\rangle$
has been computed in Appendix A and the result is
\beq
\langle e^{b\phi}e^{b\phi}\rangle=\frac{b^{-1}}{4\pi}\left(\frac{1}{b^2}-1\right)
\left[\mu\pi\gamma\left(\frac{bQ}{2}\right)\right]^{\frac{1}{b^2}-1}
\gamma\left(\frac{bQ}{2}\right)\gamma\left(\frac{1}{2}\left(1-\frac{1}{b^2}\right)\right).
\label{twob}
\eeq
As $b\to 1$, the parameter $r$ can be computed from Eqs.(\ref{descenonept}), (\ref{oneb}),
and (\ref{twob}) as follows:
\beqa
&&r=-\lim_{b\to 1}\frac{Z_{\hat{c}=1}\eta\langle e^{b\phi}\rangle}
{\sqrt{-\langle e^{b\phi}e^{b\phi}\rangle}},
\label{r2}\\
&&\langle e^{b\phi}\rangle
=-C\pi\left[\mu\pi\gamma\left(\frac{bQ}{2}\right)\right]^{\frac{1}{2}(\frac{1}{b^2}-1)}
\frac{1}{\Gamma(\frac{1}{2}(1-b^2))},
\label{oneblimit}\\
&&\langle e^{b\phi}e^{b\phi}\rangle
=-\frac{1}{2\pi}\left[\mu\pi\gamma\left(\frac{bQ}{2}\right)\right]^{\frac{1}{b^2}-1}
\frac{1}{\Gamma(\frac{1}{2}(1-b^2))}.
\label{twoblimit}
\eeqa

To compute ${\hat c}=1$ partition function $Z_{\hat{c}=1}$, it is helpful to split this
into $c=1$ free boson and $c=1/2$ fermion.
The free boson part with Neumann boundary condition is given by
$Z_{c=1}=\sqrt{R/2}$ \cite{AKK}.
The $c=1/2$ free fermion system is equivalent to a minimal CFT for the critical Ising model.
Its disk partition function $Z_{c=1/2}$ is the $p=3$ case of
\beq
Z_{r,s}=\left(\frac{8}{p(p+1)}\right)^{\frac{1}{4}}
\frac{\sin\frac{\pi r}{p}\sin\frac{\pi s}{p+1}}
{(\sin\frac{\pi}{p}\sin\frac{\pi}{p+1})^{\frac{1}{2}}},
\label{minimal}
\eeq
where $(r,s)$ denotes a conformal boundary condition \cite{Cardy}.
In this case we choose the Ising model boundary condition $(r,s)=(1,1)$, which gives
$Z_{c=1/2}=Z_{1,1}=1/\sqrt{2}$.
Substituting these into Eq.(\ref{r2}), we obtain
\beq
r=C\eta\frac{\pi^{\frac{3}{2}}\sqrt{R}}{\sqrt{2\log\frac{\Lambda}{\mu}}}.
\label{r3}
\eeq
This result agrees with the matrix model result (\ref{rmatrix}) if we can fix the
proportional constant to $C\eta=-\frac{2}{\sqrt{\pi}}$.
Since the proportional constant $C$ can be fixed only when it is compared with the matrix
model calculation, it is interesting to check if the same $C$ can be obtained
for different matrix models.
We will comment on this in the conclusion but for now we just fix the value as it is.

Now let us move on to the next order in $\la$.
We rewrite the $N=2$ perturbation term (\ref{perturbation}) as
\beq
{\cal T}={{\cal L}_{II}\over{2i\la}}=
\left[c_1\xi\bar{\xi}
-ic_2(\psi\bar{\xi}-\xi\bar{\psi})+c_3\psi\bar{\psi}\right]
\cos\left(\frac{R}{2}\tilde{x}\right)e^{(1-\frac{R}{2})\phi}
\label{calT}
\eeq
where $c_i$'s are defined in Eq.(\ref{defcis}).
From Eq.(\ref{LFTrhoparam}) the parameter $\rho$ is given by
\beq
\rho=\frac{\partial_{\lambda}Z_{disk}}{\sqrt{-\partial_{\lambda}^2F_0}}\Bigg|_{\lambda=0}
=\frac{i\langle{\cal T}\rangle_{disk}}{\sqrt{\langle{\cal T}{\cal T}\rangle_{sphere}}}.
\label{rho1}
\eeq
The one-point function $\langle{\cal T}\rangle$ in the numerator can be
computed as
\beq
\langle{\cal T}\rangle=\langle\cos\left(\frac{R}{2}\tilde{x}\right)\rangle
\left(c_1\langle\xi\bar{\xi}\rangle\langle V_{1-\frac{R}{2}}\rangle
+c_3\langle\psi\bar{\psi}V_{1-\frac{R}{2}}\rangle\right).
\label{T}
\eeq
Similarly the two-point function $\langle{\cal T}{\cal T}\rangle$ in the denominator
can be written as
\beqa
\langle{\cal T}{\cal T}\rangle&=&\langle\cos^2\left(\frac{R}{2}\tilde{x}\right)\rangle
\Bigg(-c_1^2\langle\xi\xi\rangle\langle\bar{\xi}\bar{\xi}\rangle
\langle V_{1-\frac{R}{2}}V_{1-\frac{R}{2}}\rangle)
+c_2^2\langle\bar{\xi}\bar{\xi}\rangle
\langle\psi V_{1-\frac{R}{2}}\psi V_{1-\frac{R}{2}}\rangle
\nonumber\\
&+&c_2^2\langle\xi\xi\rangle
\langle\bar{\psi}V_{1-\frac{R}{2}}\bar{\psi}V_{1-\frac{R}{2}}\rangle
+c_3^2\langle\psi\bar{\psi}V_{1-\frac{R}{2}}
\psi\bar{\psi}V_{1-\frac{R}{2}}\rangle\Bigg).
\label{TT}
\eeqa

The one-point function $\langle\xi\bar{\xi}\rangle$ in Eq.(\ref{T})
corresponds to that of the energy operator $\epsilon$ in the Ising model:
$\langle\epsilon\rangle=i\langle\xi\bar{\xi}\rangle$.
The boundary states of the Ising model are well-known
and the one-point function $\langle\epsilon\rangle$ is $\eta$.
The correlation functions of the descendant fields are related to
those of the primary field by
\beqa
&&\langle\psi\bar{\psi}V_{1-\frac{R}{2}}\rangle=i\eta\frac{1+R/2}{1-R/2}
\langle V_{1-\frac{R}{2}}\rangle,
\nonumber\\
&&\langle\psi V_{1-\frac{R}{2}}\psi V_{1-\frac{R}{2}}\rangle=\frac{1+R/2}{1-R/2}
\langle V_{1-\frac{R}{2}}V_{1-\frac{R}{2}}\rangle,
\nonumber\\
&&\langle\bar{\psi}V_{1-\frac{R}{2}}\bar{\psi}V_{1-\frac{R}{2}}\rangle=\frac{1+R/2}{1-R/2}
\langle V_{1-\frac{R}{2}}V_{1-\frac{R}{2}}\rangle,
\nonumber\\
&&\langle\psi\bar{\psi}V_{1-\frac{R}{2}}\psi\bar{\psi}V_{1-\frac{R}{2}}\rangle
=-\left(\frac{1+R/2}{1-R/2}\right)^2\langle V_{1-\frac{R}{2}}V_{1-\frac{R}{2}}\rangle.
\nonumber
\eeqa

Using these relations, $\rho$ can be rewritten as
\beq
\rho=-\sqrt{2}B\eta\frac{\langle V_{1-\frac{R}{2}}\rangle}
{\sqrt{-\langle V_{1-\frac{R}{2}}V_{1-\frac{R}{2}}\rangle}},
\label{rho2}
\eeq
where $B$ is the one point function of $\cos(\frac{R}{2}\tilde{x})$ on the disk
with Neumann boundary conditions,
which is known to be the same as $Z_{c=1}$.
The one-point function $\langle V_{b-\frac{R}{2}}\rangle$ can be derived from Eq.(\ref{one})
\beq
\langle V_{b-\frac{R}{2}}\rangle=
-C\pi\left[\mu\pi\gamma\left(\frac{bQ}{2}\right)\right]^{\frac{1}{2}
(\frac{1}{b^2}-1)+\frac{R}{2b}}
\left[b\Gamma(\frac{1}{2}(1-b^2)+\frac{bR}{2})\Gamma
(\frac{1}{2}(\frac{1}{b^2}+1)+\frac{R}{2b})\right]^{-1}.
\label{oneb-R}
\eeq

The two-point function $\langle V_{b-\frac{R}{2}}V_{b-\frac{R}{2}}\rangle$
can be computed similarly as before and its derivation is explained in details
in Appendix A.
The result is
\beqa
\langle V_{b-\frac{R}{2}}V_{b-\frac{R}{2}}\rangle
&=&\frac{b^{-1}}{4\pi}\left(\frac{1}{b^2}-1+\frac{R}{b}\right)
\left[\mu\pi\gamma\left(\frac{bQ}{2}\right)\right]^{\frac{1}{b^2}-1+\frac{R}{b}}
\gamma\left(\frac{1}{2}(1+b^2-bR)\right)
\nonumber\\
&\times&\gamma\left(\frac{1}{2}\left(1-\frac{1}{b^2}-\frac{R}{b}\right)\right).
\label{twob-R2}
\eeqa

In the limit $b\to 1$,
the one- and two-point functions approach
\beqa
&&\langle V_{b-\frac{R}{2}}\rangle
=-C\pi\left[\mu\pi\gamma\left(\frac{bQ}{2}\right)\right]^{\frac{1}{2}(\frac{1}{b^2}-1)
+\frac{R}{2b}}\frac{1}{\Gamma(R/2)\Gamma(1+R/2)},
\label{oneb-Rlimit}\\
&&\langle V_{b-\frac{R}{2}}V_{b-\frac{R}{2}}\rangle=-\frac{1}{\pi R}
\left[\mu\pi\gamma\left(\frac{bQ}{2}\right)\right]^{\frac{1}{b^2}-1+\frac{R}{b}}
\left(\frac{\Gamma(1-R/2)}{\Gamma(R/2)}\right)^2.
\label{twob-Rlimit}
\eeqa
Substituting these into (\ref{rho2}),
we obtain
\beq
\rho=\pi^{3/2}C\eta\frac{R}{\Gamma(1+R/2)\Gamma(1-R/2)}
=2C\eta\sqrt{\pi}\sin\frac{\pi R}{2},
\label{rho3}
\eeq
and the ratio
\begin{equation}
\frac{\rho}{r} = \frac {2}{\pi}\sqrt{\frac{2}{R}\log\frac\Lambda\mu}\sin\frac{\pi R}{2}.
\end{equation}
which confirms the matrix model result Eq.(\ref{ratiomatrix}).

\subsection{Two-point functions of the $N=2$ SLFT}

An interesting check of our conjecture is to compute the free energy in the large $\lambda$
limit, Eq.(\ref{freeenergyatRhalf}), where the underlying theory is the $N=2$ SLFT.
In this limit, the free energy can be expanded as
\begin{equation}
F_0(\lambda,\mu)=\sum_{n=0}^{\infty}{\mu^n\over{n!}}\bigg\langle
\left[i\int d^2z d^2\theta e^{\Phi}\right]^n\bigg\rangle_{N=2}.
\end{equation}
Here $\langle\ldots\rangle_{N=2}$ means that the correlation function is evaluated
in the context of the $N=2$ SLFT.

It is very difficult to derive general $n$-point correlation functions.
But three-point correlation functions for some special operators are computable
as we have listed in Appendix B.
We need the following three-point functions of the $N=2$ SLFT:
\beqa
\langle e^{\hat{b}(\phi^++\phi^-)}e^{\hat{b}(\phi^++\phi^-)}
\psi^-\bar{\psi}^-e^{\hat{b}\phi^+}\rangle
&=&\langle e^{\hat{b}(\phi^++\phi^-)}e^{\hat{b}(\phi^++\phi^-)}\psi^+\bar{\psi}^+e^{\hat{b}
\phi^-}\rangle
\nonumber\\
&=&\lambda^{\frac{2}{\hat{b}^2}-5}\frac{\hat{b}^{-13}}{\sqrt{2}}
\frac{\gamma(1-\hat{b}^2)^2\gamma(2-1/\hat{b}^2)(2\hat{b}^2-1)^2}{\gamma(1-2\hat{b}^2)}.
\label{e+-e+-e+2i}
\eeqa
Integrating these and using the relation (\ref{twon1}),
we obtain
\beq
\langle\psi\bar{\psi}e^{\hat{b}(\phi^++\phi^-)}\psi\bar{\psi}e^{\hat{b}(\phi^++\phi^-)}\rangle
=-\frac{\sqrt{2}}{2}i\hat{b}^{-9}
(\frac{1}{\hat{b}^2}-1)^2(2\hat{b}^2-1)\frac{\gamma(1-\hat{b}^2)^2\gamma(2-1/\hat{b}^2)}
{\gamma(1-2\hat{b}^2)}\lambda^{\frac{2}{\hat{b}^2}-4}.
\label{e+-e+-i}
\eeq
When we fix $\hat{b}=1/\sqrt{2}$, the two-point function vanishes.

Noticing that $\psi{\bar\psi}e^{\phi}$ is the $N=1$ SLFT interaction, we can expect that
the small $\mu$ corrections to the $N=2$ SLFT free energy should not have $\mu^2$ terms
which can be checked easily in the matrix model result (\ref{freeenergyatRhalf}).
In fact, this result shows that general $n$-point ($n\ge 2$) functions of the $N=1$ SLFT
interaction terms vanish.
But confirming this in the $N=2$ SLFT context is beyond the scope of this paper.

\section{Conclusion}

In this paper we have proposed a world sheet action corresponding to
the deformed type 0A matrix model which describes 2D fermionic black hole.
The model is the $N=1$ SLFT perturbed by the $N=2$ SLFT interaction.
To justify our conjecture, we have computed the non-perturbative effects arising from the
D-instanton for the $N=1$ SLFT with $N=2$ perturbation and its proposed dual matrix model
and have shown the agreement between the two results.
This confirms our conjecture for the $N=1$ SLFT side.
When the $N=2$ SLFT interaction becomes strong, we have calculated the correlation
function of two $N=1$ SLFT interaction operators and confirmed that both the $N=2$ SLFT
and the deformed matrix model agree.
Although these calculations are not mathematical proof of the duality conjecture,
we believe that they are enough to justify the validity.

For non-perturbative corrections, we have computed the $r$ parameter which measures
the D-instanton effect of the $N=1$ SLFT without the $N=2$ SLFT interaction and
the $\rho$ parameter which is the first order $N=2$ SLFT perturbation effect.
Since the dual relation between the $N=1$ SLFT and
the type 0A matrix model is well established \cite{newhat}, we certainly expect that
$r$ should match at both sides. Assuming this, our computation shows
that $\rho$ matches at both sides, which confirms the same
D-instanton effect in the perturbed case. In order to decide the
precise value of $r$ we should fix the undetermined quantity $C\eta$ in
the computation. This cannot be done solely in the $N=1$ SLFT setup.
Following the approach of \cite{AKK}, one can
consider the $N=1$ SLFT coupled to superminimal models and decide
the similar constant $C\eta$ by comparing with the corresponding matrix
models and see if we can get the same $C\eta$ for various superminimal
models and the ${\hat c}=1$ model.
Our preliminary analysis shows a discrepancy by a factor 2 which we can
not explain yet.
We may need a better
understanding of the string equation governing superminimal models
and to work out the non-perturbative effects of such theories as in
\cite{ZE}. This in itself is an interesting topic and we hope we can
report the progress elsewhere but is beyond the scope of the Liouville
theory calculation, which is of our main interest in this paper.

If this duality is accepted, one can compare the free energy of the deformed matrix model
(\ref{Fpmzero}) with the small $\lambda$ expansion of the SLFT free energy.
This leads to a supersymmetric version of G. Moore's conjecture for the general
$n$-point correlation functions on the sphere of the certain $N=1$ SLFT operators:
\begin{equation}
\bigg\langle {{\cal T}_R}^n ({\cal T}_R^*)^n\bigg\rangle_{0}=
2Rn!\mu^2\left[(1-2R)\mu^{R-2}\right]^n{\Gamma(n(2-2R)-2)\over{\Gamma(n(1-2R)+1)}}
\end{equation}
where
\begin{equation}
{\cal T}_R=i\int d^2 zd^2\theta  e^{i\frac{R}{2}\tilde{X}+(1-\frac{R}{2})\Phi}.
\end{equation}
Due to technical difficulties of the $N=1$ SLFT, this conjecture can not be proved
by direct calculations.
But this result seems natural because the ordinary Liouville theory and the $N=1$ SLFT
share many physical properties like coupling constant duality in common.

\section*{Acknowledgement}

This work was supported by KOSEF-M60501000025-05A0100-02500 (CA,MY),
KRF-2004-042-C00023 (JP,MY) and JSPS (TS).

\section*{Appendix A: Two-point functions of the $N=1$ SLFT}

The three-point function of the $N=1$ SLFT is given by \cite{FH}
\beqa
&&-\alpha_1^2\langle\psi\bar{\psi}N_{\alpha_1}N_{\alpha_2}N_{\alpha_3}\rangle
\nonumber\\
&&=i\left[\mu\pi\gamma\left(\frac{bQ}{2}\right)b^{1-b^2}\right]^{\frac{Q-\sum\alpha_i}{b}}
\frac{2\Upsilon'_{NS}(0)\Upsilon_{NS}(2\alpha_1)
\Upsilon_{NS}(2\alpha_2)\Upsilon_{NS}(2\alpha_3)}
{\Upsilon_R(\alpha_{1+2+3}-Q)\Upsilon_R(\alpha_{1+2-3})
\Upsilon_R(\alpha_{2+3-1})\Upsilon_R(\alpha_{3+1-2})},
\label{three}
\eeqa
where
\beq
\Upsilon_{NS}(x)=\Upsilon\left(\frac{x}{2}\right)\Upsilon\left(\frac{x+Q}{2}\right),
~~~\Upsilon_R(x)=\Upsilon\left(\frac{x+b}{2}\right)\Upsilon\left(\frac{x+b^{-1}}{2}\right),
\label{GammaNSR}
\eeq
and $\Upsilon(x)$ satisfies the following formulae
\beqa
&&\Upsilon'(0)=\Upsilon(b),
\nonumber\\
&&\Upsilon(x+b)=\gamma(bx)b^{1-2bx}\Upsilon(x),
\nonumber\\
&&\Upsilon(x+1/b)=\gamma(x/b)b^{2x/b-1}\Upsilon(x).
\nonumber
\eeqa
For $\alpha_1=\alpha_2=\alpha_3=b$,
using the formulae
\beqa
&&\Upsilon'_{NS}(0)=\frac{1}{2}\Upsilon_R(b),
\nonumber\\
&&\frac{\Upsilon_{NS}(2b)}{\Upsilon_R(b)}
=\gamma\left(\frac{bQ}{2}\right)b^{1-bQ},
\nonumber\\
&&\frac{\Upsilon_{NS}(2b)}{\Upsilon_R(2b-\frac{1}{b})}
=-\gamma\left(\frac{1}{2}\left(1-\frac{1}{b^2}\right)\right)
\left(\frac{bQ}{2}-1\right)^2b^{-2-\frac{1}{b^2}},
\nonumber
\eeqa
we find
\beq
-b^2\langle\psi\bar{\psi}N_bN_bN_b\rangle
=-i(\mu\pi)^{\frac{1}{b^2}-2}\left[\gamma\left(\frac{bQ}{2}\right)\right]^{\frac{1}{b^2}}b^{-5}
\gamma\left(\frac{1}{2}\left(1-\frac{1}{b^2}\right)\right)\left(\frac{bQ}{2}-1\right)^2.
\label{threeb2}
\eeq
Integrating this,
we obtain
\beq
\langle e^{b\phi}e^{b\phi}\rangle=\frac{b^{-1}}{4\pi}\left(\frac{1}{b^2}-1\right)
\left[\mu\pi\gamma\left(\frac{bQ}{2}\right)\right]^{\frac{1}{b^2}-1}
\gamma\left(\frac{bQ}{2}\right)\gamma\left(\frac{1}{2}\left(1-\frac{1}{b^2}\right)\right).
\label{twobi}
\eeq

For the two-point function $\langle V_{b-\frac{R}{2}}V_{b-\frac{R}{2}}\rangle$,
one can integrate a three-point function
$\langle\psi\bar{\psi}V_bV_{b-\frac{R}{2}}V_{b-\frac{R}{2}}\rangle$.
Setting $\alpha_1=b$, $\alpha_2=\alpha_3=b-\frac{R}{2}$ in (\ref{three})
and using the formulae
\beqa
&&\frac{\Upsilon_{NS}(2b-R)}{\Upsilon_R(b-R)}
=\gamma\left(b\left(\frac{Q}{2}-\frac{R}{2}\right)\right)b^{1+bR-bQ},
\nonumber\\
&&\frac{\Upsilon_{NS}(2b-R)}{\Upsilon_R(2b-\frac{1}{b}-R)}
=-\gamma\left(\frac{1}{2}\left(1-\frac{1}{b^2}-\frac{R}{b}\right)\right)
\left(b\left(\frac{Q}{2}-\frac{R}{2}\right)-1\right)^2b^{-2-\frac{1}{b^2}-\frac{R}{b}},
\nonumber
\eeqa
we find
\beqa
-b^2\langle\psi\bar{\psi}V_bV_{b-\frac{R}{2}}V_{b-\frac{R}{2}}\rangle
&=&-i\left[\mu\pi\gamma\left(\frac{bQ}{2}\right)\right]^{\frac{1}{b^2}-2+\frac{R}{b}}
\gamma\left(\frac{bQ}{2}\right)b^{-5}
\gamma\left(b\left(\frac{Q}{2}-\frac{R}{2}\right)\right)
\nonumber\\
&\times&\gamma\left(\frac{1}{2}\left(1-\frac{1}{b^2}-\frac{R}{b}\right)\right)
\left(b\left(\frac{Q}{2}-\frac{R}{2}\right)-1\right)^2.
\label{threeb-R2}
\eeqa
Integrating this,
we obtain
\beqa
\langle V_{b-\frac{R}{2}}V_{b-\frac{R}{2}}\rangle
&=&\frac{b^{-1}}{4\pi}\left(\frac{1}{b^2}-1+\frac{R}{b}\right)
\left[\mu\pi\gamma\left(\frac{bQ}{2}\right)\right]^{\frac{1}{b^2}-1+\frac{R}{b}}
\gamma\left(\frac{1}{2}(1+b^2-bR)\right)
\nonumber\\
&\times&\gamma\left(\frac{1}{2}\left(1-\frac{1}{b^2}-\frac{R}{b}\right)\right).
\label{twob-R2}
\eeqa

\section*{Appendix B: Two-point functions of the $N=2$ SLFT}

We consider the vertex operators of the following form
\beq
V^{j(s,\bar{s})}_{m,\bar{m}}=\exp\left[\sqrt{\frac{2}{k}}(i(m+s)x_L+i(\bar{m}+\bar{s})x_R-j\phi)
+isH_L+i\bar{s}H_R\right],
\label{vertex}
\eeq
where $H_{L,R}$ are bosonized fermions
and $s, \bar{s}\in Z$ for the NS sector,
$s, \bar{s}\in Z+\frac{1}{2}$ for the R sector.
The constant $k$ is related to the $N=2$ SLFT coupling constant $\hat{b}$ by $k=\hat{b}^2$.
The dimensions are given by
\beq
\Delta=-\frac{j(j+1)}{k}+\frac{(m+s)^2}{k}+\frac{s^2}{2},
~~~\bar{\Delta}=-\frac{j(j+1)}{k}+\frac{(\bar{m}+\bar{s})^2}{k}+\frac{\bar{s}^2}{2},
\label{dimension}
\eeq
and the U(1) charges are given by
\beq
\omega=\frac{2(m+s)}{k}+s,
~~~\bar{\omega}=\frac{2(\bar{m}+\bar{s})}{k}+\bar{s}.
\label{charge}
\eeq
When $s=\bar{s}=0$,
$m$ and $\bar{m}$ are given by
\beq
m=\frac{n+wk}{2},
~~~\bar{m}=\frac{n-wk}{2},
\label{mm}
\eeq
where $n$ and $w$ are momentum and winding numbers respectively.

The two-point function of the operator
$\psi\bar{\psi}e^{\sqrt{2}\hat{b}\phi}=\psi\bar{\psi}e^{\hat{b}(\phi^++\phi^-)}$
satisfies the following relation
\beq
\langle\psi\bar{\psi}e^{\hat{b}(\phi^++\phi^-)}\psi\bar{\psi}e^{\hat{b}(\phi^++\phi^-)}\rangle
=-(\frac{1}{\hat{b}^2}-1)^2\langle e^{\hat{b}(\phi^++\phi^-)}e^{\hat{b}(\phi^++\phi^-)}\rangle.
\label{twon1}
\eeq
The operator $e^{\hat{b}(\phi^++\phi^-)}$ is identified with the vertex operator of the form (\ref{vertex}):
$e^{\hat{b}(\phi^++\phi^-)}=V^{-k(0,0)}_{0,0}$.
The two-point function $\langle e^{\hat{b}(\phi^++\phi^-)}e^{\hat{b}(\phi^++\phi^-)}\rangle$
is computed by integrating the three-point function
$\langle e^{\hat{b}(\phi^++\phi^-)}e^{\hat{b}(\phi^++\phi^-)}\psi^-\bar{\psi}^-e^{\hat{b}\phi^+}\rangle$
and $\langle e^{\hat{b}(\phi^++\phi^-)}e^{\hat{b}(\phi^++\phi^-)}\psi^+\bar{\psi}^+e^{\hat{b}\phi^-}\rangle$
with respect to $-i\hat{b}^2\lambda$.
The three-point function of $N=2$ SLFT is given in \cite{Hosomichi}.
Using the result there,
we find
\beq
\langle e^{\hat{b}(\phi^++\phi^-)}e^{\hat{b}(\phi^++\phi^-)}\psi^-\bar{\psi}^-e^{\hat{b}\phi^+}\rangle
=\langle V^{-k(0,0)}_{0,0}V^{-k(0,0)}_{0,0}V^{-k/2(-1,-1)}_{k/2+1,k/2+1}\rangle
=F_-D_-.
\label{e+-e+-e+1}
\eeq
The expressions of $F_-$ and $D_-$ are written in \cite{Hosomichi}
and they are calculated as
\beqa
F_-&=&-\frac{1}{\hat{b}^4}\gamma(1-\hat{b}^2)^3,
\nonumber\\
D_-&=&\nu^{-\frac{5}{2}\hat{b}^2+1}\frac{\hat{b}^{6\hat{b}^2+2/\hat{b}^2-6}}{\sqrt{2}}
\frac{\Upsilon(-2\hat{b}+1/\hat{b})^2\Upsilon(-\hat{b}+1/\hat{b})}
{\Upsilon(3\hat{b}-1/\hat{b})\Upsilon(\hat{b})\Upsilon(2\hat{b})}
\nonumber\\
&=&-\lambda^{\frac{2}{\hat{b}^2}-5}\frac{\hat{b}^{-9}}{\sqrt{2}}
\frac{\gamma(2-1/\hat{b}^2)(2\hat{b}^2-1)^2}{\gamma(1-2\hat{b}^2)\gamma(1-\hat{b}^2)},
\nonumber
\eeqa
and we obtain
\beq
\langle e^{\hat{b}(\phi^++\phi^-)}e^{\hat{b}(\phi^++\phi^-)}\psi^-\bar{\psi}^-e^{\hat{b}\phi^+}\rangle
=\lambda^{\frac{2}{\hat{b}^2}-5}\frac{\hat{b}^{-13}}{\sqrt{2}}
\frac{\gamma(1-\hat{b}^2)^2\gamma(2-1/\hat{b}^2)(2\hat{b}^2-1)^2}{\gamma(1-2\hat{b}^2)}.
\label{e+-e+-e+2}
\eeq
Similarly we find
\beqa
\langle e^{\hat{b}(\phi^++\phi^-)}e^{\hat{b}(\phi^++\phi^-)}\psi^+\bar{\psi}^+e^{\hat{b}\phi^-}\rangle
&=&\langle V^{-k(0,0)}_{0,0}V^{-k(0,0)}_{0,0}V^{-k/2(1,1)}_{-k/2-1,-k/2-1}\rangle
\nonumber\\
&=&\langle e^{\hat{b}(\phi^++\phi^-)}e^{\hat{b}(\phi^++\phi^-)}\psi^-\bar{\psi}^-e^{\hat{b}\phi^+}\rangle.
\label{e+-e+-e-}
\eeqa
Integrating the sum of (\ref{e+-e+-e+2}) and (\ref{e+-e+-e-}),
we obtain
\beq
\langle e^{\hat{b}(\phi^++\phi^-)}e^{\hat{b}(\phi^++\phi^-)}\rangle
=\frac{\sqrt{2}}{2}i\hat{b}^{-9}(2\hat{b}^2-1)\frac{\gamma(1-\hat{b}^2)^2\gamma(2-1/\hat{b}^2)}
{\gamma(1-2\hat{b}^2)}\lambda^{\frac{2}{\hat{b}^2}-4}.
\label{e+-e+-}
\eeq
Substituting in (\ref{twon1}),
we obtain (\ref{e+-e+-i}).

\end{document}